\documentclass[aps,prl,twocolumn,showpacs,superscriptaddress,preprintnumbers,amsmath,amssymb]{revtex4-1}

\usepackage{amssymb}
\usepackage{color}
\usepackage{graphicx}
\usepackage{epstopdf}
\usepackage[hypertexnames=false,colorlinks=true,linkcolor=blue,citecolor=blue]{hyperref}

\graphicspath{{eps/}{pdf/}}
\newcommand{\R}{\mathbb{R}}

\def\Re{\mathop{\mathrm{Re}}}

\def\Xint#1{\mathchoice
   {\XXint\displaystyle\textstyle{#1}}%
   {\XXint\textstyle\scriptstyle{#1}}%
   {\XXint\scriptstyle\scriptscriptstyle{#1}}%
   {\XXint\scriptscriptstyle\scriptscriptstyle{#1}}%
   \!\int}
\def\XXint#1#2#3{{\setbox0=\hbox{$#1{#2#3}{\int}$}
     \vcenter{\hbox{$#2#3$}}\kern-.5\wd0}}

\newcommand{\K}{{K}}

\def\dashint{\Xint-}

\newcommand{\rmO}{\mathrm{O}}

\newcommand{\rmd}{\mathrm{d}}
\newcommand{\rme}{\mathrm{e}}
\newcommand{\rmi}{\mathrm{i}}

\begin{document}

% \preprint{APS}
\title{Universal wavenumber selection laws in apical growth}
\author{Ryan Goh}
\author{Rajendra Beekie}
\affiliation{School of Mathematics, University of Minnesota, Minneapolis, MN 55455}
\author{Daniel Matthias}
\affiliation{Department of Applied Mathematics, University of Colorado, Boulder, CO 80305}
\author{Joshua Nunley}
\affiliation{Department of Mathematica Sciences,
University of Arkansas,
Fayetteville, AR 72701}
\author{Arnd Scheel}
\affiliation{School of Mathematics, University of Minnesota,  Minneapolis, MN 55455}

\date{\today}

% \preprint{considered for publication in Phys. Rev. E}

 %82.40.Ck	, 
%87.18.Hf 	Spatiotemporal pattern formation in cellular populations

%45.70.Qj Pattern formation
%02.30.Oz Math. methods: Bifurcation theory
%68.03.Kn fluid interfaces

%47.54.-r Fluid dynamics: Pattern selection; pattern formation

%47.11 KB Computational Methods in Fluid Dynamics: spectral methods (???)
%05.65.+b Self-organized systems 
%89.75.Kd Interdisciplinary application to physics: Patterns
%05.45.-a Nonlinear dynamics and chaos

\begin{abstract}
We study pattern-forming dissipative systems in growing domains. We characterize classes of boundary conditions that allow for defect-free growth and derive universal scaling laws for the wavenumber in the bulk of the domain. 
Scalings are based on a description of striped patterns in semi-bounded domains via strain-displacement relations. We compare predictions with direct simulations in the Swift-Hohenberg, the Complex Ginzburg-Landau, the Cahn-Hilliard, and reaction-diffusion equations.
\end{abstract}

\maketitle

% \begin{acknowledgments}
% The authors acknowledge partial support through NSF-DMS-1311740.
% \end{acknowledgments}

%something more on universality in the introduction?
Pattern-forming systems such as  fluid convection problems, reaction-diffusion systems near Turing instabilities, diblock copolymers, or phase separation problems often  exhibit striped phases, that is, stable or metastable periodic structures with wavenumbers in an admissible band $k\in (k_-,k_+)$. In large aspect-ratio systems, one typically sees a mixture of patches evolve from random initial conditions with different wavenumbers, that may be separated by defects or mix slowly via diffusive repair \cite{mix1,mix2}. On the other hand, it has long been known that growth processes tend to select specific wavenumbers $k_\mathrm{gr}$ from the admissible band, leading to perfect, defect-free periodic structures \cite{maini,bradley,phyllo,thiele,pattern1}. Such growth of periodic structures is fairly well understood when patterns grow by spreading into an unstable state \cite{invasion,invasion2}, in the wake of a free invasion front with speed $c_\mathrm{free}$. Here, we are interested in situations when growth is externally imposed. We therefore consider systems on time-depending domains $x\in [-L(t),L(t)]$, or with a parameter $\mu(x,t)$ that drives pattern formation in $[ -L(t),L(t)]$. 

For growth speeds $L'(t)\equiv c\gg 1$, one often observes a spatially 
homogeneous equilibrium state near the boundary that is subsequently invaded by 
a pattern-forming front with speed $c_\mathrm{free}$. For $L'(t)\equiv c\lesssim 
c_\mathrm{free}$, the patterns selected are close to patterns selected by the 
free invasion front \cite{goh1,goh2}. Our aim here is to  derive asymptotic 
expressions for the selected wavenumber $k=k(c)$ when  $c\ll 1$, applicable to 
a variety of pattern-forming systems.

\begin{figure}[h]
	\centering
	\includegraphics[width=0.49\linewidth]{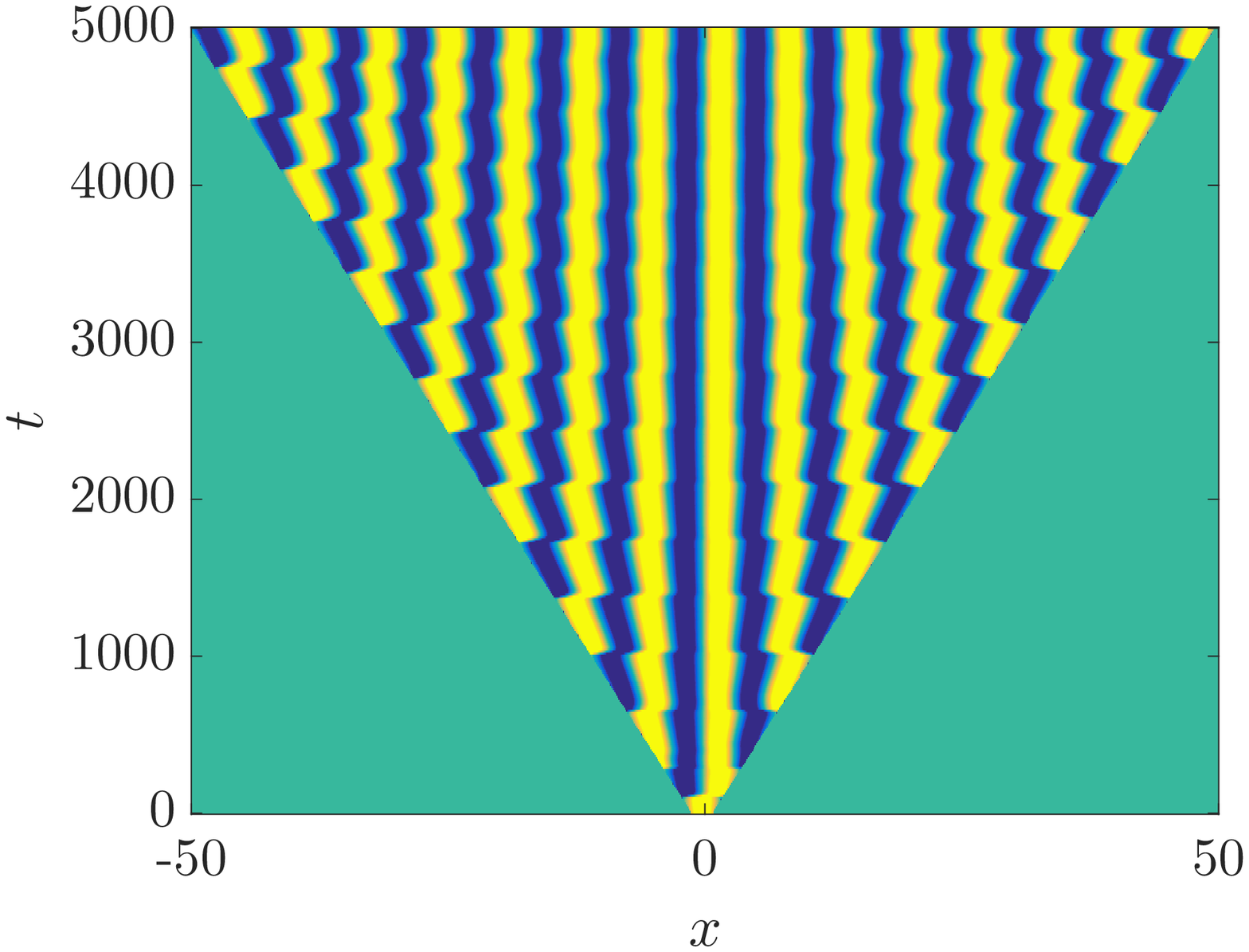}\ 
	\includegraphics[width=0.49\linewidth]{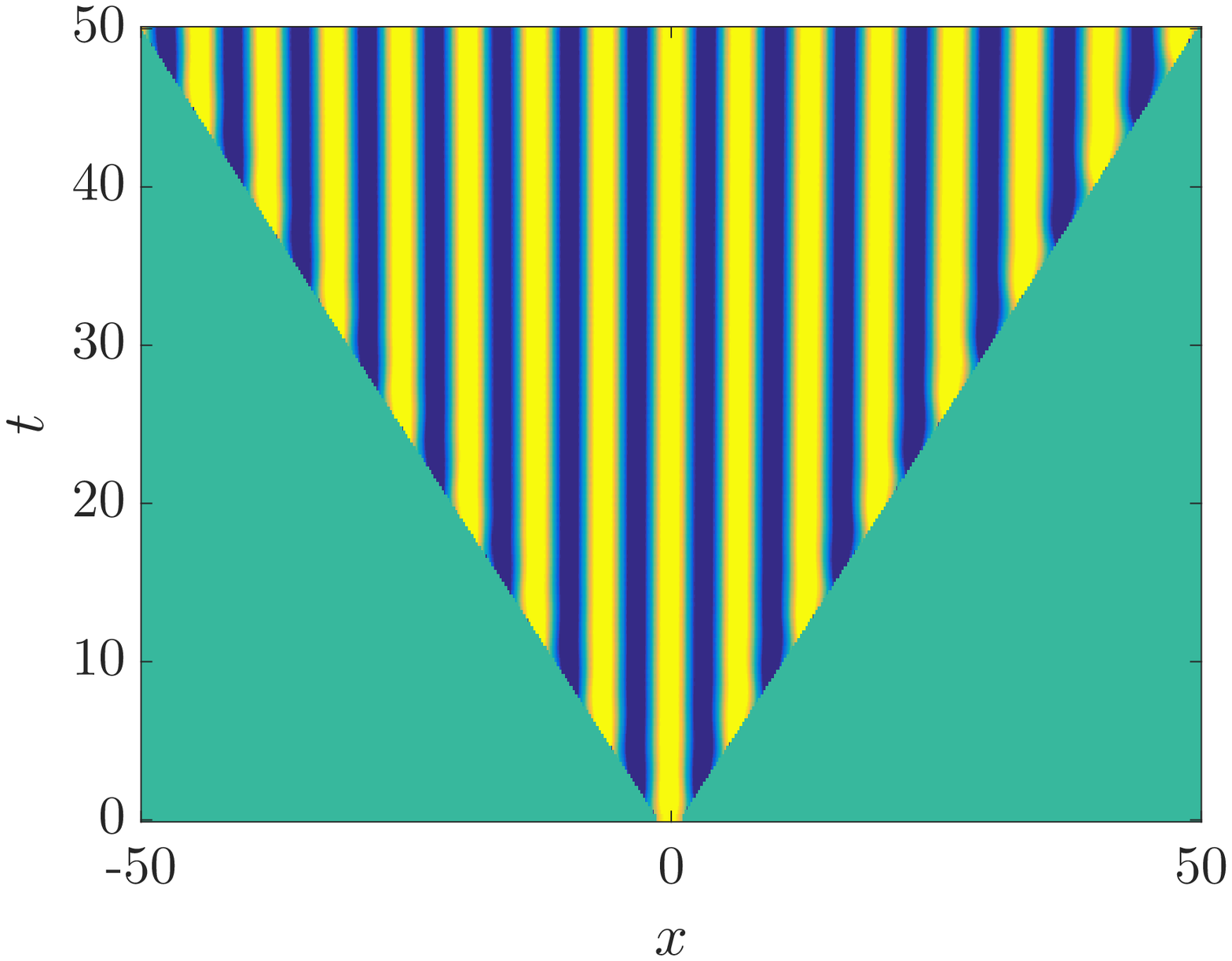}
% * V-shaped plot of growing domain pattern, maybe two speeds; take figure from Josh (check!)
% 
	\vspace*{-.25in}
\caption{Apical growth in SH \eqref{e:sh} with $\mu = 1.5$%--\eqref{e:shbc}
, $c=0.01$ (left, $k\sim 0.928$) and $c=1$ (right, $k\sim 0.981$).}\label{f:2}
\end{figure}

\begin{figure}[h]
	\centering
% 	\includegraphics[width=1.0\linewidth]{}
% * spatial dynamics picture
% * actual plot of W^s, BC in CGL?
% * strain-displacement curve, insets for actual patterns
		\includegraphics[width=0.39\linewidth]{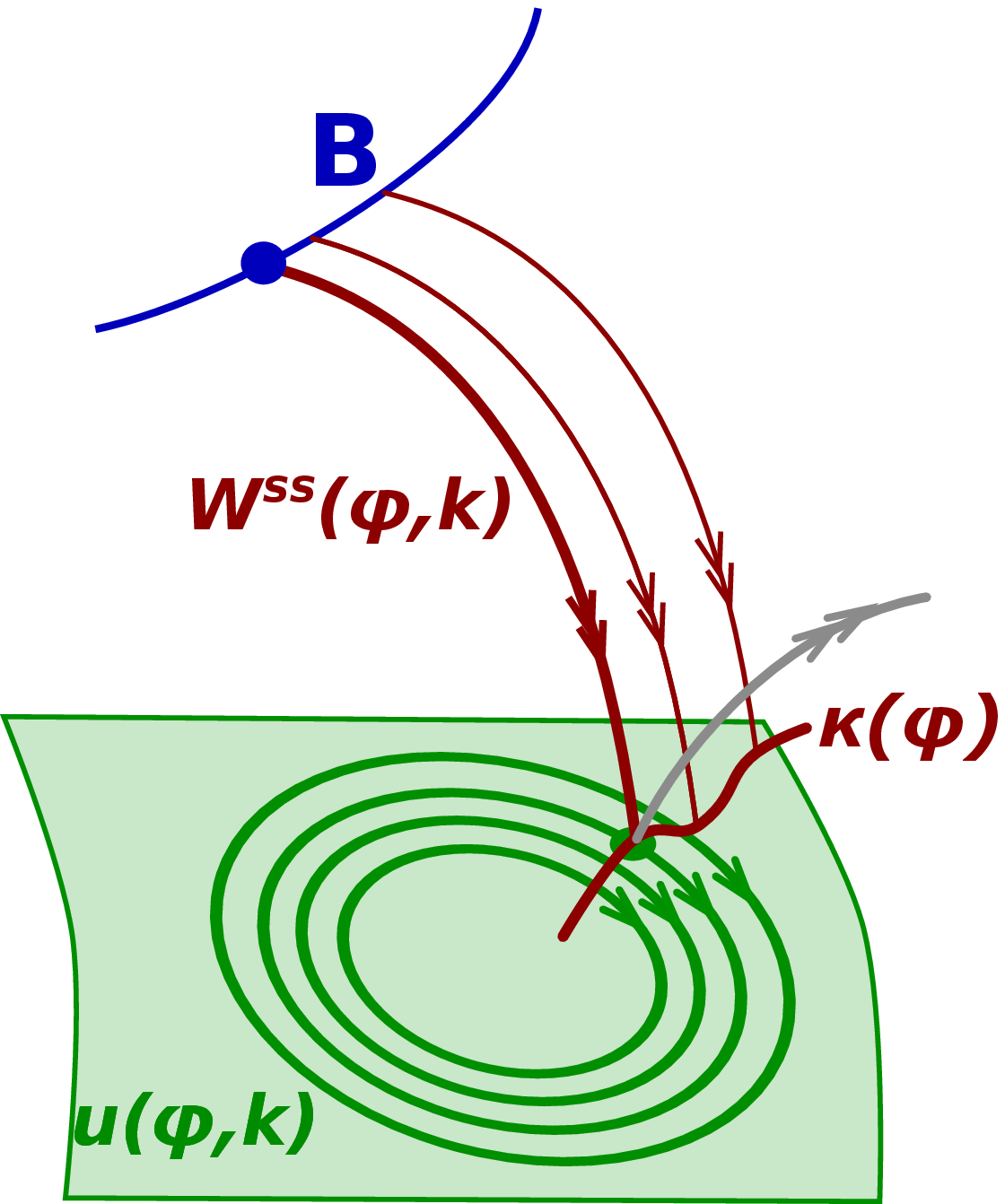}
		\includegraphics[width=0.59\linewidth]{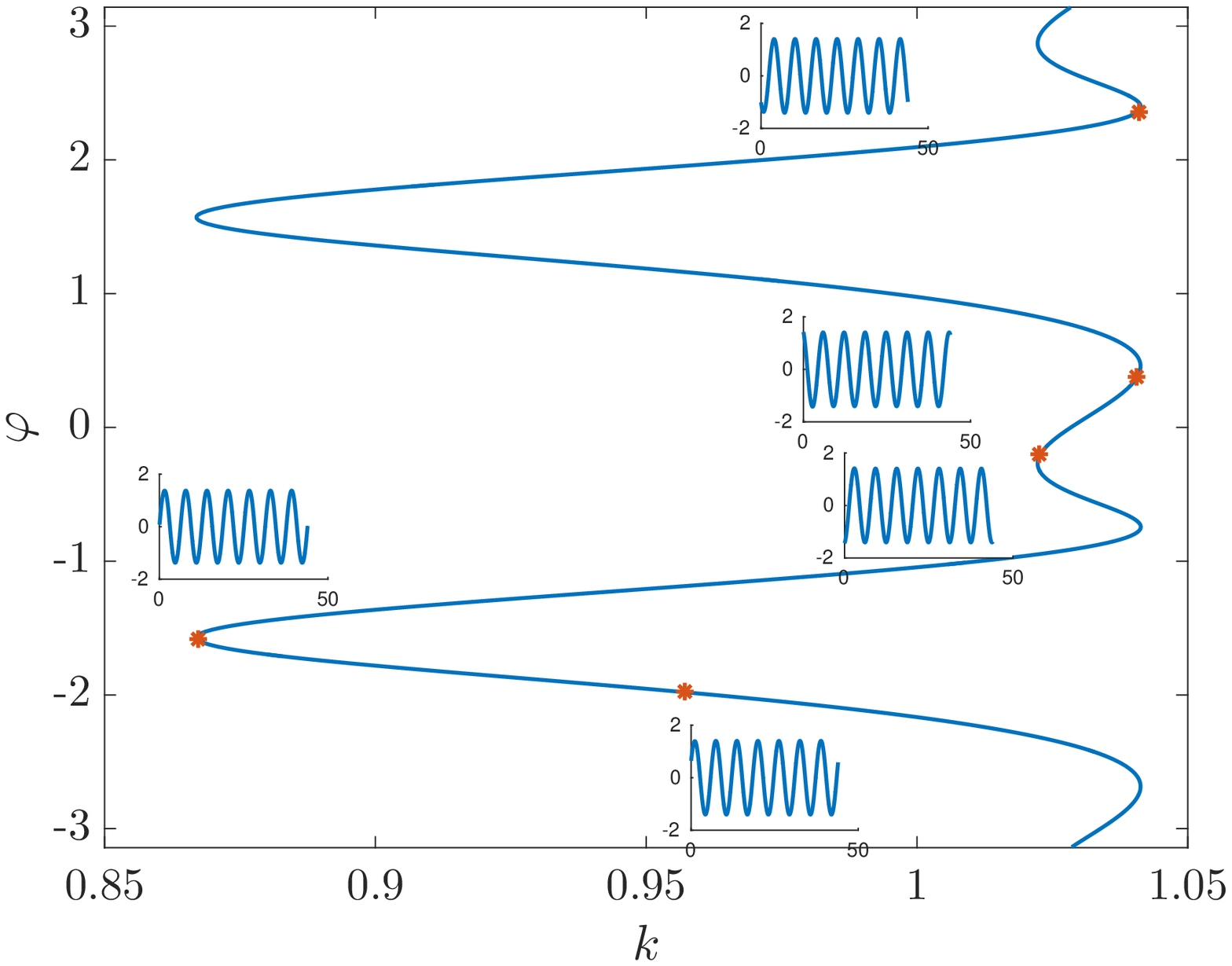}
	\caption{Schematic illustration of the shooting problem, connecting boundary conditions to periodic orbits (left); strain-displacement curve of \eqref{e:sh} with free b.c. and $\mu=1.5$, select boundary layer profiles as insets in the figure (right).}\label{f:1}
\end{figure}

\begin{figure}[h]
	\centering
\includegraphics[width=1.0\linewidth]{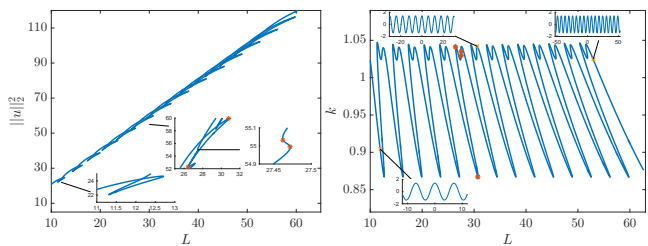}
	\caption{Equilibria of \eqref{e:sh} %--\eqref{e:shbc}
	continued in the domain size $L$, with select saddle-nodes marked as orange stars; saddle-nodes approach extrema of $\K $ \eqref{e:tp}. }\label{f:sn}
\end{figure}

The Swift-Hohenberg equation 
\begin{equation}\label{e:sh}
u_t=-(\partial_{xx}+1)^2u + \mu u - u^3,
\end{equation}
is a prototypical example for the formation of striped patterns. For fixed 
$\mu$, there exists a family of periodic, even solutions, parameterized by the 
wavnumber, $u_\mathrm{st}(kx;k)$, 
$u_\mathrm{st}(\xi+2\pi;k)=u_\mathrm{st}(\xi;k)=u_\mathrm{st}(-\xi;k)$. We 
consider \eqref{e:sh}  with ``free'' boundary conditions 
\begin{equation}\label{e:shbc}
u_{xx}+u=(u_{xx}+u)_x=0,
\end{equation}
induced by the $L^2$-gradient flow to the free energy 
\begin{equation}\label{e:en}
E(u)=\int\left( (u_{xx}+u)^2- \mu u^2+\frac{1}{2}u^4\right)\rmd x.
\end{equation}
Direct simulations in a growing domain show a dependence of the wavenumber in 
the bulk of the domain on the speed of growth. For slow speeds, the dynamics 
near the edge are governed by long transients where patterns ``lock'' to the 
boundary, separated by sudden snapping where the phase at the boundary 
jumps (Fig. \ref{f:2}).   
Neglecting the effect of the second, far-away boundary, we consider \eqref{e:sh} on $x\in(-ct,\infty)$. Seeking to perturb from $c=0$, we start with the description of ``boundary layer'' type equilibria of \eqref{e:sh} on the half line  $x\in (0,\infty)$ that satisfy boundary conditions and that are asymptotic to periodic solutions. Such equilibria arise as intersections of the 2d-subspace in  4d-phase space $(u,u_x,u_{xx},u_{xxx})$ that satisfies the boundary conditions, with the 3d-stable manifold of periodic solutions. One therefore expects equilibria to occur in one-parameter families $u_*(x;\tau)$,
\[
\lim_{x\to\infty}|u_*(x;\tau)-u_\mathrm{st}(k(\tau)x-\varphi(\tau));k(\tau)|=0.
\]
Following boundary layers in the parameter $\tau$, one notices, far away from the boundary,  variations in wavenumber (strain) and an effective phase shift (displacement) relative to the boundary (Fig. \ref{f:1}). We therefore refer to the curves $k(\tau),\varphi(\tau)$ as strain-displacement (SD) relations \cite{mor}.   SD relations can be computed explicitly at small amplitudes, exploiting integrability of amplitude equations, and numerically at finite amplitude using numerical continuation \cite{mor}. 
For $\mu$ not too large, SD-relations turn out to be wavenumber selecting, $k=\K (\varphi)\in (k_-,k_+)$, within the Eckhaus-stable band. 
At minima and maxima of $\K $, boundary layers undergo a saddle-node bifurcation and branches with $\K '>0$ are stable. Equilibria in bounded domains can be readily constructed from displacement-strain relations by imposing a simple phase- and wavenumber matching in the center of the domain with exponentially small corrections from the boundary layers. Restricting, for simplicity, to even solutions, we may impose Neumann boundary conditions at $x=0$, which gives  
\begin{equation}\label{e:pm}
\K (\varphi)L=\varphi\mod 2\pi.
\end{equation}
The wavenumber-selecting SD-relations associated with free boundary conditions then yields  a snaking bifurcation diagram in the domain size $L$ (Fig. \ref{f:sn}).
Adiabatic growth, that is, letting the system relax to equilibrium each time after increasing the domain size, induces snapping as observed in Fig. \ref{f:2} near the turning point values $k_\mathrm{tp}$. These can be computed from phase matching \eqref{e:pm} solving $\rmd \varphi/\rmd L=0$, which gives  $\K '(\varphi)=\K (\varphi)/\varphi=1/L$,  and for large $L$
\begin{equation}\label{e:tp}
k_\mathrm{tp}=\K(\varphi_\mathrm{min})+\frac{1}{4\K _2} L^{-2}+\ldots,
\end{equation}
where we used an expansion $\K(\varphi)=\K(\varphi_\mathrm{min})+\K_2 (\varphi-\varphi_\mathrm{min})^2+\ldots$
Implementing adiabatic growth numerically or experimentally therefore allows one to directly  measure $\K (\varphi)$ on stable branches of SD-relations. 
Growth at constant speed is however non-adiabatic, since the relaxation to equilibrium is diffusive in large domains, eventually slower than the linear growth. 

Based on SD relations, we now derive an asymptotic formula for $k(c)$ in the case of constant rate growth $c=L'(t)$. Since patterns during the growth process are well approximated by boundary layers near the minimum $\varphi\sim\varphi_\mathrm{min}$, leading-order expansions can be derived from a phase-diffusion approximation with effective diffusivity $d_\mathrm{eff}$ evaluated at $k_\mathrm{min}=\K(\varphi_\mathrm{min})$,
\begin{equation}\label{e:pd2}
\left\{\begin{array}{rll}
\vartheta_t&=d_\mathrm{eff} \vartheta_{xx} -c\vartheta_x,& \quad x>0;\\ \vartheta_x&=\K (\vartheta), &\quad x=0,
\end{array}\right.
\end{equation}
where effective boundary conditions are induced by the strain-displacement relation \footnote{Note that only wavenumber-selecting strain-displacement relations $k=K(\varphi)$ yield well-posed mixed boundary conditions \cite{mor}.}. 
% 
% They therefore yield well-posed boundary conditions in a phase-diffusion approximation,
% \begin{equation}\label{e:pd}
% \left\{\begin{array}{rll}
% \vartheta_t&=(d(\vartheta_x))_x,& \quad x>0;\\ \vartheta_x&=\K (\vartheta), &\quad x=0.
% \end{array}\right.
% \end{equation}
% where $d(k)$ denotes the effective diffusivity at wavenumber $k$.  Note however that the multiple-scale derivation used to justify \eqref{e:pd} is usually not valid near the boundary. 
The growth process is described by time-periodic solutions to \eqref{e:pd2} with linear asymptotics,
%\[
$
\vartheta(t,x)=\vartheta(t+\frac{2\pi}{\omega},x)$, $\vartheta(t,x)\sim kx \mbox{ for }x\to\infty, 
$
%\]
$\omega=ck$. Substituting $\vartheta=\theta+kx-\omega t$ gives
\begin{equation}\label{e:pd3}
\left\{\begin{array}{rll}
\theta_t&=d_\mathrm{eff} \theta_{xx} -c\theta_x,& \quad x>0;\\ \theta_x&=\K (\theta-\omega t)-k, &\quad x=0.
\end{array}\right.
\end{equation}
Requiring pinning of the phase at the boundary except at snapping points implies  $\theta-\omega t\equiv \theta_\mathrm{min} \mod 2\pi$ at the boundary. Neglecting the higher-order term $c\theta_x$ yields
\begin{equation}\label{e:pd4}
\left\{\begin{array}{rll}
\theta_t&=d_\mathrm{eff} \theta_{xx},& \quad x>0;\\ \theta&=\omega t+\theta_\mathrm{min} \mod 2\pi, &\quad x=0,
\end{array}\right.
\end{equation}
with explicit, leading-order outer solution 
\[
\theta_\mathrm{out}(t,x)=\theta_\mathrm{min}+\sum_{\ell\neq 0}(-1)^\ell (\rmi \ell)^{-1} \rme^{\rmi \ell \omega t-\sqrt{\rmi \ell\omega/d_\mathrm{eff}}\; x},
\]
where we used the branch cut $\Re(\sqrt{\rmi \ell\omega/d_\mathrm{eff}})>0$. 
Substituting into the boundary conditions of \eqref{e:pd3} shows that the approximation by $\theta_\mathrm{out}$ holds until snapping, when  $\partial_x\theta_\mathrm{out}(t_\mathrm{snap},0)=k_\mathrm{min}-k$. For periodicity, we require $t_\mathrm{snap}=2\pi/(ck)$, and obtain
\begin{equation}\label{e:jump}
k-k_\mathrm{min}=-\partial_x\theta_\mathrm{out}(\frac{2\pi}{ck},0)\sim -\zeta(\frac{1}{2})\sqrt{2} c_\mathrm{pe}^{1/2},
\end{equation}
where $\zeta$ is the Riemann $\zeta$-function, $\zeta(\frac{1}{2})\sim-1.460$, and  $c_\mathrm{pe}=ck_\mathrm{min}/d_\mathrm{eff}$ is a non-dimensionalized speed similar to a P\'eclet number.  The $\zeta$-function arises through the limit $\mathrm{Re}\{\lim_{t\nearrow(2\pi/ck)} \partial_x\theta_\mathrm{out}(t,0)\}$ which is obtained from the analytic continuation to $s = 1/2$ of the polylogarithms $L_s(z) = \sum_{n=1}^\infty\frac{(z)^n}{(n)^s}$
% ,\,\tilde L_s(z) = \sum_{n=1}^\infty\frac{(z)^{-n}}{(-n)^s}$, 
as $z$ approaches 1 along the unit-circle counter-clockwise.
%another line or so?
The next order of the expansion is determined by the passage through the minimum of $\K $. Expanding $\K (\theta)=k_\mathrm{min}+\K _2 \theta^2+\ldots$, we find 
\begin{equation}\label{e:riccati}
\left\{\begin{array}{rll}
\theta_t&=d_\mathrm{eff} \theta_{xx},& \quad x>0;\\ \theta_x&=\K _2\theta^2-\partial_{tx}\theta_\mathrm{out}|_{t=2\pi/ck}\cdot t, &\quad x=0,
\end{array}\right.
\end{equation}
for the next order. Scaling yields a Riccati-type flux,
\begin{equation}\label{e:riccati2}
%\left\{\begin{array}{rll}
%\tilde{\theta}_\tau&=d_\mathrm{eff} \tilde{\theta}_{yy},& \quad x>0;\\ \tilde{\theta}_y&=\K _2\tilde{\theta}^2+\tau, &\quad x=0,
%\end{array}\right.
\left\{\begin{array}{rll}
\tilde{\theta}_\tau&=\tilde{\theta}_{yy},& \quad y>0;\\ \tilde{\theta}_y&=\tilde{\theta}^2+\tau, &\quad y=0.
\end{array}\right.
\end{equation}
% Solutions exhibit boundary  blowup at some finite time $\tau_\mathrm{sn}$, which determines the time instance when snapping occurs. 
%
Unlike the analysis of a slow passage through a saddle-node, where the blowup time in the Riccati equations uniquely determines the bifurcation delay,  solutions with $\theta|_{x=0}\sim \sqrt{-\tau}$ for $\tau\to -\infty$ here come in a one-parameter family.  Solutions in this family exhibit boundary blowup at times  $\tau_\mathrm{sn}\in(2,7)$. Compatibility with the periodicity $\omega=ck$ then gives the expansion 
%\begin{equation}\label{e:asy}
%k(c) = k_\mathrm{min}   -\zeta(\frac{1}{2})(2 c_\mathrm{pe})^{1/2}+\sqrt{2}\,\zeta(-\frac{1}{2})\tau_\mathrm{sn} \K _2^{-\frac{1}{2}}c_\mathrm{pe}^{3/4}+\ldots.
%% \sqrt{\frac{2c k_\mathrm{min} }{d}}\left[ -\zeta(\frac{1}{2}) + \zeta(-\frac{1}{2}) \tau_\mathrm{sn} \K _2^{-\frac{1}{2}} \left( \frac{c k_\mathrm{min}}{d_\mathrm{eff}}  \right)^{1/4}  \right]
%\end{equation}
% \begin{equation}\label{e:asy}
% k(c) = k_\mathrm{min}   -\,\zeta(\frac{1}{2})(2 c_\mathrm{pe})^{\frac{1}{2}}  - 
% \,2^{\frac{1}{4}}\zeta(-\frac{1}{2})^{\frac{1}{2}}\tau_\mathrm{sn} \K 
% _2^{-\frac{1}{2}}c_\mathrm{pe}^{\frac{3}{4}}+\ldots.
% % \sqrt{\frac{2c k_\mathrm{min} }{d}}\left[ -\zeta(\frac{1}{2}) + 
% % \zeta(-\frac{1}{2}) \tau_\mathrm{sn} \K _2^{-\frac{1}{2}} \left( \frac{c 
% % k_\mathrm{min}}{d_\mathrm{eff}}  \right)^{1/4}  \right]
% \end{equation}
% \begin{equation}\label{e:asy}
% k(c) = k_\mathrm{min}   
% -2^{\frac{1}{2}}\,\zeta(\frac{1}{2}) c_\mathrm{pe}^{\frac{1}{2}}  - 
% \,2^{\frac{1}{4}}\zeta(-\frac{1}{2})^{\frac{1}{2}}\tau_\mathrm{sn} \K 
% _2^{-\frac{1}{2}}c_\mathrm{pe}^{\frac{3}{4}}+\ldots.
% \end{equation}
\begin{equation}\label{e:asy}
% \begin{array}{c}
k(c) = k_\mathrm{min} + k_{1/2}c_\mathrm{pe}^{1/2} +k_{3/4}c_\mathrm{pe}^{3/4}+\ldots,
% \end{array}
\end{equation}
\[
k_{1/2}=-2^{1/2}\,\zeta(1/2),\ 
k_{3/4}= - \,2^{1/4}|\zeta(-1/2))|^{1/2}\tau_\mathrm{sn} \K_2^{-1/2}.
\]
% 
% -2^{\frac{1}{2}}\,\zeta(\frac{1}{2}) c_\mathrm{pe}^{\frac{1}{2}}  - 
% \,2^{\frac{1}{4}}\zeta(-\frac{1}{2})^{\frac{1}{2}}\tau_\mathrm{sn} \K 
% _2^{-\frac{1}{2}}c_\mathrm{pe}^{\frac{3}{4}}+\ldots.
% \end{equation}
% 
% \begin{itemize}
% \item $k_\mathrm{min}$ - minimum of strain displacement curve
% \item $d_\mathrm{eff}$ - Effective diffusion
% \item $\zeta(x)$ - Riemann Zeta function 
% \item $\tilde\tau_\mathrm{sn}$ - Ricatti constant,  i.e. time for passage through fold
% \item $\K _2$ - Coefficient of 2nd order term in taylor jet, i.e. half of the 2nd derivative of strain displacement curve at $k_\mathrm{min}$.
% \end{itemize}
The snapping itself is described by a global heteroclinic orbit connecting $\theta(x)\equiv\theta_\mathrm{min}$ to $\theta(x)=\theta_\mathrm{min}-2\pi$ in \eqref{e:pd3} with $c =0$. Converting the heat equation into a boundary integral equation, 
\[
\theta(t)=\int_0^t(\pi(t-s))^{-\frac{1}{2}}\left(\K (\theta(s))-k_\mathrm{min}-\theta_0(s)\right)\rmd s,
\]
where $\theta_0$ accounts for initial conditions,
one finds a fractional differential equation with saddle-node equilibrium $\theta_\mathrm{min}$. Exploiting monotonicity and asymptotics near the saddle-node \cite{olmstead}, one can readily establish the existence of such a heteroclinic in this  case of the phase-diffusion equation.
% 
% %\[
% %\left\{\begin{array}{rll}
% %\theta_t&=d_\mathrm{eff} \theta_{xx},& \quad x>0;\\ \theta_x&=\K (\theta)-k_\mathrm{min}, &\quad x=0,
% %\end{array}\right.
% %\]
% Since the flux is quadratic near its minimum, the equilibria are of saddle-node type and the heteroclinic is a saddle-node heteroclinic in the fractional differential equation that one obtains after converting the heat equation into a boundary intergral equation 
% \[
% \theta=\int_0^t(\pi(t-s))^{-\frac{1}{2}}\left(\K (\theta)-k_\mathrm{min}-\theta_0(s)\right)\rmd s.
% \]
% %sqrt pi, initial conditions, or better formulate as infinite integral
% Using asymptotics near the saddle-node equilibria \cite{olmstead} and monotonicity, one can readily establish the existence of such a heteroclinic in this  case of the phase-diffusion equation. 
We emphasize however that the global heteroclinic solution is not universally described by the phase-diffusion approximation since it occurs on an $\rmO(1)$ time-scale. Indeed, we  observed in the Complex Ginzburg-Landau equation that snapping events may at times involve nucleation of defects. 

Note that the transition from stationary boundary layers, $c=0$, to periodic nucleation with large period $\sim 1/c$ cannot be viewed as a saddle-node bifurcation on a limit cycle which would in fact predict $T\sim 1/\sqrt{c}$. 

\begin{figure}[h]
	\centering
	\includegraphics[width=1.0\linewidth]{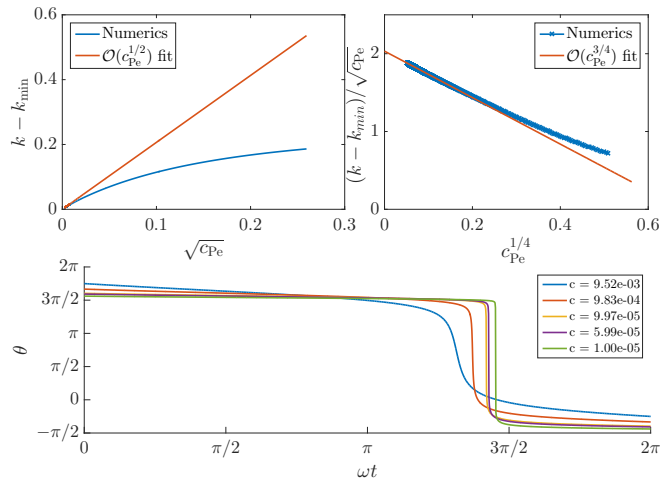}
	\caption{Asymptotically linear relation of $k-k_\mathrm{min}$ vs $\sqrt{c_\mathrm{pe}}$ with slope $\sqrt{2}\zeta(1/2)$ (top left); plot exhibiting 1/4 exponent in corrections (top right); sample plots of $\theta(t,0)$ for range of $c$ values (bottom);   $K (\theta)=1+0.3\sin(\theta)$, $d_\mathrm{eff}=1.$}\label{f:falcon}
% PICTURES:
% * illustrate main result with FALCON computations; two examples? 
% * log plots showing accuracy of higher-order coefficients
%       (k-k_min)/sqrt(c) against c^1/4 plots are basically linear!
% * show picture k-m_min vs sqrt(c) to exhibit relevance of next-order term    
% * time series at boundary (without subtracting omega t)
% * space-time plot of solution (without overlay of omega t?)
\end{figure}

We corroborated the asymptotics \eqref{e:asy} numerically (Fig. \ref{f:falcon}), converting \eqref{e:pd3} into a boundary integral equation
\begin{equation}\label{e:pd3f}
D_{c,k}\theta=\K (\theta-\omega t)-k,
\end{equation}
with pseudo-differential operator $D_{c,k}$ defined by its Fourier multiplier 
$
\hat{D}_{c,k}(\ell)=
\left(1-\sqrt{1-4d_\mathrm{eff}\rmi ck\ell}\right)/2d_\mathrm{eff}.
$
Adding a phase condition $\dashint\theta=0$ with associated Lagrange multiplier $k$, we used pseudo-arclength continuation to continue periodic solutions in $c$ down to $c=10^{-5}$, using $2^{17}$ Fourier modes for various SD-relations. Boundary profiles show the characteristic snapping behavior. Extrapolating a plot of $k$ vs $\sqrt{c_\mathrm{pe}}$ gives intercept $k_\mathrm{min}$ and slope $k_{1/2}\approx 2.0653$ within $10^{-2}$ accuracy. We fitted the next-order coefficient to values  $\tau_\mathrm{sn}\in [2, 7]$ in good agreement with direct computations of the blowup times in \eqref{e:riccati2} and obtained improved approximations of the asymptotic expansions. 

%Results for different choices of  SD-relations $\K(\theta) $ match the expansions equally well. 

\begin{figure}[h]
	\centering
 	\includegraphics[width=1.0\linewidth]{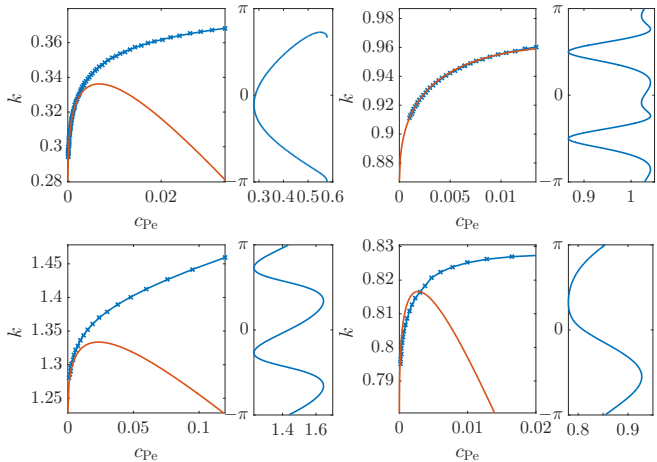}
%\vspace*{1.5in}
	\caption{Measured (blue crosses) vs predicted speeds (orange line) with 
associated strain-displacement relations  for (top left to bottom right) CGL 
\eqref{e:cgl}, ($\mu_k=0.4,\mu_0=0.1, d_\mathrm{eff} \approx  0.830$), SH 
\eqref{e:sh} ($\mu=1.5,d_\mathrm{eff} \approx  3.531 $),  RD \eqref{e:rd} 
($\mu_0 = 0.95, d_u = 0.1, d_v = 2,\gamma = 0.2,d_\mathrm{eff} \approx 0.872$), 
and CH \eqref{e:ch} ($\mu_k=0.6,\mu_\theta=0.15,\mu_2=0.2, d_\mathrm{eff} 
\approx 0.8275$).  Errors in the leading order coefficient  $k_{1/2}=-\sqrt{2}\zeta(1/2)$ are $0.0061$, $-0.0033$, $0.0914$, and $-0.1970$ 
respectively.}\label{f:univ}
% Raj's notes say mu_theta=1.5 in one place and 0.15 in another --- check!
	
% * PICTURES:
% * several graphs:
%   * overlay of prediction and measurement-- absolute scales, one graph k-kmin vs c, plot of theoretical curves, data points, 
%   

\end{figure}

Since the leading-order expansions can be derived from slow variations near a 
phase $\theta=\theta_\mathrm{min}$, one expects the asymptotics to be 
universally valid. We tested our predictions in several pattern-forming systems. 
We first considered the Swift-Hohenberg equation \eqref{e:sh} with free boundary 
conditions and $\mu=1.5$.  In order to obtain predictions from \eqref{e:asy}, we computed
strain-displacement relations and effective diffusivities numerically \cite{mor}.
% 
% In direct simulations, we used the method of lines with second-order finite 
% differences $dx=0.005$ and adaptive time-stepping on fixed domains of size 
% $L=500,100,2000$ for decreasing speeds $c$, imposing artificial boundary 
% conditions at $x=L$. Simulations at small speeds require very long relaxation 
% times and large domains due to weakly exponentially decay effects from the 
% arbitrary right boundary. 
%
Fig. \ref{f:univ}  compares asymptotics and data from direct 
simulations \footnote{In direct simulations, for CGL, SH, and RD, we used domain sizes $L= 500\ldots 2000$, for CH $L=2000
\ldots 8000$. We used second order finite differences with $dx=0.005\ldots 0.1$ for spatial discretization, and Matlab's ODE15s for time stepping.}.
%CGL - $L = 500,...,2000 dx = 0.1$
%SH -  $L = 500,...,2000, dx = 0.005$
%RD - $L = 2000, dx = 0.06$
%CH - $L = 500,....8000, dx = 0.1$
 We note that $k_\mathrm{min}<k_\mathrm{zz}$, the zigzag boundary, that is, patterns formed in slow growth processes are stretched 
relative to the energy-minimizing equilibrium strain. 
We also compared results for the Complex Ginzburg-Landau equation (Fig. \eqref{f:univ}), 
\begin{equation}\label{e:cgl}
\left\{\begin{array}{rll}
A_t&=A_{xx}+A-A|A|^2,& \quad x>0;\\ A_x&=\mu_k\rmi  A + \mu_0, &\quad x=0.
\end{array}\right.
\end{equation}
SD-relations can be computed explicitly since the steady-state equation $0=A_{xx}+A-A|A|^2$ is integrable \cite{mor}. 
% We remark that the SD-curve in this case is only locally wavenumber selecting near $k_\mathrm{min}$ as the curve enters the Eckhaus-unstable region for large $k$. %We found similar phenomena for a range of boundary conditions $A_x= \alpha A +\beta$. %Don't need this sentence???

%For $|\Im\alpha|$ small, SD-relations allow for $k=0$ and growth selects $k=0$. Increasing $|\Im\alpha|$, SD-relations develop a tangency with $k=0$ and selected wavenumbers detach from $k=0$ with the universal square root asymptotics of a saddle-node bifurcation on a limit cycle, which translates into
%\[
%k=2\sqrt{\K_2}\sqrt{k_\mathrm{min}}+\ldots
%\]
Instead of imposing boundary conditions at $x=-ct$, one  can also envision situations when a parameter $\mu=\mu(x+ct)$  allows for periodic patterns when $\xi=x+ct>0$, large, but possesses a trivial stable state when $\xi<0$. We explored such situations in \eqref{e:sh} and in an activator-inhibitor  reaction-diffusion system 
\begin{equation}
\left\{\begin{array}{rll}
u_t&=d_uu_{xx}+\mu(x-ct)u - u^3 - v \\
v_t&=d_vv_{xx}+u - \gamma v,\,\,\,\, x\in\R,%mu = \pm0.95, D_u = 0.1, D_v =2; 
\end{array}\right.\label{e:rd}
\end{equation}
where $\mu(\xi) = \pm \mu_0$ for $\xi\gtrless0$.
The convergence to a trivial state as $\xi\to -\infty$ imposes an effective boundary condition on patterns in $\xi>0$, for which one can compute SD-relations at $c=0$. For $|\mu_0|$ not too large, SD-relations select wavenumbers and one encounters similar asymptotics for small speeds $c$  (Fig. \ref{f:univ}). Our last example is the (integrated) Cahn-Hilliard equation
\begin{equation}\label{e:ch}
\left\{\begin{array}{rll}
\theta_t&=-(\theta_{xxx}+\theta_x-\theta_x^3)_x,& \quad x>0;\\ \theta_x&=\mu_k+\mu_\theta \sin(\theta),\ \  \theta_{xx}=\mu_2, &\quad x=0,
\end{array}\right.
\end{equation}
Again, SD-relations can be computed explicitly and are wavenumber selecting for a large class of parameters $\mu_k,\mu_\theta,\mu_2$ (Fig \ref{f:univ}). 
We note that our study here is confined to wavenumber-selecting SD-relations. Decreasing $\mu_k$ in CH, one can explore limitations: SD-relations touch the Eckhaus boundary and one observes nucleation of kink defects. 
% 
% Neumann boundary conditions, or free boundary conditions with $\mu>0.7$ achieve the minimum at the Eckhaus boundary $k_\mathrm{min}=k_\mathrm{eck}$ and growth will be accompanied by defect nucleation near the boundary. 
\begin{figure}[h]
	\centering
 	\includegraphics[width=1.0\linewidth]{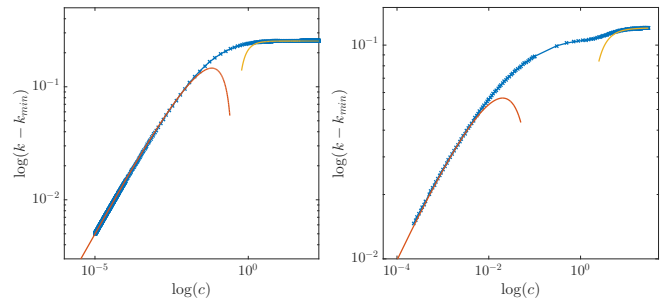}
\caption{Selected wavenumbers plotted on a log scale for small and large speeds in phase-diffusion \eqref{e:pd2}, $K(\theta)=1+0.3\sin(\theta)$, and in CGL \eqref{e:cgl}. Blue curves show direct simulations, orange and gold curves show small and large speed predictions.}\label{f:large}
\end{figure}

In the phase-diffusion approximation, for large speeds, one can neglect diffusion to find $\vartheta_t=c\vartheta_x$, reducing the problem on the boundary to an ODE with wavenumber given as the harmonic average of $\K $,
\[
\vartheta_t=c \K (\vartheta), \quad  k_\mathrm{h}=\left(\dashint \K (\vartheta)^{-1}\rmd\vartheta\right)^{-1}.
\]
At next order, one finds $k(c)=k_\mathrm{h}-k_2c^{-2}$, $k_2=\dashint ((\K')^2/K) (\dashint \K^{-2})^{-2}.$
% include next order?
% For intermediate and large speeds, the phase-diffusion approximation is generally invalid and one does not expect universal laws. Nevertheless, within the approximation, for l
Those asymptotics are not universal since the modulation approximation breaks down at intermediate speeds. Nevertheless, similar asymptotics in CGL \eqref{e:cgl}, give $k=\mu_k + \mu_0^2\mu_k^{-5}\left(\frac{7}{2}\mu_0^2-2\mu_k^2+2\mu_k^4\right)c^{-2}\sim 0.4-0.228 c^{-2}$ for $\mu_k=0.4,\mu_0=0.1$; see Fig. \ref{f:large} for comparisons.
% wavenumber measurements and predictions in PD and CGL.
% \begin{figure}[h]
% 	\centering
% % 	\includegraphics[width=1.0\linewidth]{}
% \vspace*{1.5in}
% 	\caption{Measured vs predicted speeds for examples RD \eqref{e:rd},SH \eqref{e:sh}, CGL \eqref{e:cgl}, and CH \eqref{e:ch}; associated strain-displacement relations}\label{f:univ}
% 
% % * PICTURES:
% % * several graphs:
% %   * overlay of prediction and measurement-- absolute scales, one graph k-kmin vs c, plot of theoretical curves, data points, 
% %   
% \end{figure}

Summarizing, we derived asymptotics for the wavenumber selected in the bulk of pattern forming systems through apical growth at uniform rate. The predictions are based on SD-relations, which characterize patterns in fixed, semi-infinite domains. Defect-free growth is possible for wavenumber selecting SD relations, where  $k_\mathrm{min}$ lies within the Eckhaus-stable band.
We obtained good comparison between predictions and direct numerical simulations in a variety of pattern-forming systems, including a reaction-diffusion system, the Swift-Hohenberg, Cahn-Hilliard, and Complex Ginzburg-Landau equations. SD-relations can be measured directly in experiments when growth is adiabatic. We therefore envision that our predictions would compare well with experiments such as B\'enard convection. Our approach should also give quantitative predictions for the distortion of higher-dimensional patterns, such as hexagonal lattices created in apical growth.

 \begin{acknowledgments}
 \section{Acknowledgments}
 The authors acknowledge partial support from NSF-DMS-1311740, NSF-GFRP-00006595, and a UMN DDF. \end{acknowledgments}
\vspace{-0.3in}

\bibliographystyle{apsrev4-1}
\bibliography{growing_domains_prl}

\begin{thebibliography}{15}%
\makeatletter
\providecommand \@ifxundefined [1]{%
 \@ifx{#1\undefined}
}%
\providecommand \@ifnum [1]{%
 \ifnum #1\expandafter \@firstoftwo
 \else \expandafter \@secondoftwo
 \fi
}%
\providecommand \@ifx [1]{%
 \ifx #1\expandafter \@firstoftwo
 \else \expandafter \@secondoftwo
 \fi
}%
\providecommand \natexlab [1]{#1}%
\providecommand \enquote  [1]{``#1''}%
\providecommand \bibnamefont  [1]{#1}%
\providecommand \bibfnamefont [1]{#1}%
\providecommand \citenamefont [1]{#1}%
\providecommand \href@noop [0]{\@secondoftwo}%
\providecommand \href [0]{\begingroup \@sanitize@url \@href}%
\providecommand \@href[1]{\@@startlink{#1}\@@href}%
\providecommand \@@href[1]{\endgroup#1\@@endlink}%
\providecommand \@sanitize@url [0]{\catcode `\\12\catcode `\$12\catcode
  `\&12\catcode `\#12\catcode `\^12\catcode `\_12\catcode `\%12\relax}%
\providecommand \@@startlink[1]{}%
\providecommand \@@endlink[0]{}%
\providecommand \url  [0]{\begingroup\@sanitize@url \@url }%
\providecommand \@url [1]{\endgroup\@href {#1}{\urlprefix }}%
\providecommand \urlprefix  [0]{URL }%
\providecommand \Eprint [0]{\href }%
\providecommand \doibase [0]{http://dx.doi.org/}%
\providecommand \selectlanguage [0]{\@gobble}%
\providecommand \bibinfo  [0]{\@secondoftwo}%
\providecommand \bibfield  [0]{\@secondoftwo}%
\providecommand \translation [1]{[#1]}%
\providecommand \BibitemOpen [0]{}%
\providecommand \bibitemStop [0]{}%
\providecommand \bibitemNoStop [0]{.\EOS\space}%
\providecommand \EOS [0]{\spacefactor3000\relax}%
\providecommand \BibitemShut  [1]{\csname bibitem#1\endcsname}%
\let\auto@bib@innerbib\@empty
%</preamble>
\bibitem [{\citenamefont {Collet}\ \emph {et~al.}(1992)\citenamefont {Collet},
  \citenamefont {Eckmann},\ and\ \citenamefont {Epstein}}]{mix1}%
  \BibitemOpen
  \bibfield  {author} {\bibinfo {author} {\bibfnamefont {P.}~\bibnamefont
  {Collet}}, \bibinfo {author} {\bibfnamefont {J.-P.}\ \bibnamefont {Eckmann}},
  \ and\ \bibinfo {author} {\bibfnamefont {H.}~\bibnamefont {Epstein}},\
  }\href@noop {} {\bibfield  {journal} {\bibinfo  {journal} {Helv. Phys. Acta}\
  }\textbf {\bibinfo {volume} {65}},\ \bibinfo {pages} {56} (\bibinfo {year}
  {1992})}\BibitemShut {NoStop}%
\bibitem [{\citenamefont {Gallay}\ and\ \citenamefont {Mielke}(1998)}]{mix2}%
  \BibitemOpen
  \bibfield  {author} {\bibinfo {author} {\bibfnamefont {T.}~\bibnamefont
  {Gallay}}\ and\ \bibinfo {author} {\bibfnamefont {A.}~\bibnamefont
  {Mielke}},\ }\href {\doibase 10.1007/s002200050495} {\bibfield  {journal}
  {\bibinfo  {journal} {Comm. Math. Phys.}\ }\textbf {\bibinfo {volume}
  {199}},\ \bibinfo {pages} {71} (\bibinfo {year} {1998})}\BibitemShut
  {NoStop}%
\bibitem [{\citenamefont {Crampin}\ \emph {et~al.}(1999)\citenamefont
  {Crampin}, \citenamefont {Gaffney},\ and\ \citenamefont {Maini}}]{maini}%
  \BibitemOpen
  \bibfield  {author} {\bibinfo {author} {\bibfnamefont {E.}~\bibnamefont
  {Crampin}}, \bibinfo {author} {\bibfnamefont {E.}~\bibnamefont {Gaffney}}, \
  and\ \bibinfo {author} {\bibfnamefont {P.}~\bibnamefont {Maini}},\ }\href
  {\doibase 10.1006/bulm.1999.0131} {\bibfield  {journal} {\bibinfo  {journal}
  {Bulletin of Mathematical Biology}\ }\textbf {\bibinfo {volume} {61}},\
  \bibinfo {pages} {1093} (\bibinfo {year} {1999})}\BibitemShut {NoStop}%
\bibitem [{\citenamefont {Gelfand}\ and\ \citenamefont
  {Bradley}(2012)}]{bradley}%
  \BibitemOpen
  \bibfield  {author} {\bibinfo {author} {\bibfnamefont {M.~P.}\ \bibnamefont
  {Gelfand}}\ and\ \bibinfo {author} {\bibfnamefont {R.~M.}\ \bibnamefont
  {Bradley}},\ }\href {\doibase 10.1103/PhysRevB.86.121406} {\bibfield
  {journal} {\bibinfo  {journal} {Phys. Rev. B}\ }\textbf {\bibinfo {volume}
  {86}},\ \bibinfo {pages} {121406} (\bibinfo {year} {2012})}\BibitemShut
  {NoStop}%
\bibitem [{\citenamefont {Pennybacker}\ \emph {et~al.}(2015)\citenamefont
  {Pennybacker}, \citenamefont {Shipman},\ and\ \citenamefont
  {Newell}}]{phyllo}%
  \BibitemOpen
  \bibfield  {author} {\bibinfo {author} {\bibfnamefont {M.~F.}\ \bibnamefont
  {Pennybacker}}, \bibinfo {author} {\bibfnamefont {P.~D.}\ \bibnamefont
  {Shipman}}, \ and\ \bibinfo {author} {\bibfnamefont {A.~C.}\ \bibnamefont
  {Newell}},\ }\href {\doibase http://dx.doi.org/10.1016/j.physd.2015.05.003}
  {\bibfield  {journal} {\bibinfo  {journal} {Physica D: Nonlinear Phenomena}\
  }\textbf {\bibinfo {volume} {306}},\ \bibinfo {pages} {48 } (\bibinfo {year}
  {2015})}\BibitemShut {NoStop}%
\bibitem [{\citenamefont {{Wilczek, M.}}\ \emph {et~al.}(2015)\citenamefont
  {{Wilczek, M.}}, \citenamefont {{Tewes, W. B.H.}}, \citenamefont {{Gurevich,
  S. V.}}, \citenamefont {{K\"{o}pf, M. H.}}, \citenamefont {{Chi, L. F.}},\
  and\ \citenamefont {{Thiele, U.}}}]{thiele}%
  \BibitemOpen
  \bibfield  {author} {\bibinfo {author} {\bibnamefont {{Wilczek, M.}}},
  \bibinfo {author} {\bibnamefont {{Tewes, W. B.H.}}}, \bibinfo {author}
  {\bibnamefont {{Gurevich, S. V.}}}, \bibinfo {author} {\bibnamefont
  {{K\"{o}pf, M. H.}}}, \bibinfo {author} {\bibnamefont {{Chi, L. F.}}}, \ and\
  \bibinfo {author} {\bibnamefont {{Thiele, U.}}},\ }\href {\doibase
  10.1051/mmnp/201510402} {\bibfield  {journal} {\bibinfo  {journal} {Math.
  Model. Nat. Phenom.}\ }\textbf {\bibinfo {volume} {10}},\ \bibinfo {pages}
  {44} (\bibinfo {year} {2015})}\BibitemShut {NoStop}%
\bibitem [{\citenamefont {Foard}\ and\ \citenamefont
  {Wagner}(2012)}]{pattern1}%
  \BibitemOpen
  \bibfield  {author} {\bibinfo {author} {\bibfnamefont {E.~M.}\ \bibnamefont
  {Foard}}\ and\ \bibinfo {author} {\bibfnamefont {A.~J.}\ \bibnamefont
  {Wagner}},\ }\href {\doibase 10.1103/PhysRevE.85.011501} {\bibfield
  {journal} {\bibinfo  {journal} {Phys. Rev. E}\ }\textbf {\bibinfo {volume}
  {85}},\ \bibinfo {pages} {011501} (\bibinfo {year} {2012})}\BibitemShut
  {NoStop}%
\bibitem [{\citenamefont {Dee}\ and\ \citenamefont {Langer}(1983)}]{invasion}%
  \BibitemOpen
  \bibfield  {author} {\bibinfo {author} {\bibfnamefont {G.}~\bibnamefont
  {Dee}}\ and\ \bibinfo {author} {\bibfnamefont {J.~S.}\ \bibnamefont
  {Langer}},\ }\href {\doibase 10.1103/PhysRevLett.50.383} {\bibfield
  {journal} {\bibinfo  {journal} {Phys. Rev. Lett.}\ }\textbf {\bibinfo
  {volume} {50}},\ \bibinfo {pages} {383} (\bibinfo {year} {1983})}\BibitemShut
  {NoStop}%
\bibitem [{\citenamefont {van Saarloos}(2003)}]{invasion2}%
  \BibitemOpen
  \bibfield  {author} {\bibinfo {author} {\bibfnamefont {W.}~\bibnamefont {van
  Saarloos}},\ }\href {\doibase
  http://dx.doi.org/10.1016/j.physrep.2003.08.001} {\bibfield  {journal}
  {\bibinfo  {journal} {Physics Reports}\ }\textbf {\bibinfo {volume} {386}},\
  \bibinfo {pages} {29 } (\bibinfo {year} {2003})}\BibitemShut {NoStop}%
\bibitem [{\citenamefont {Goh}\ and\ \citenamefont {Scheel}(2014)}]{goh1}%
  \BibitemOpen
  \bibfield  {author} {\bibinfo {author} {\bibfnamefont {R.}~\bibnamefont
  {Goh}}\ and\ \bibinfo {author} {\bibfnamefont {A.}~\bibnamefont {Scheel}},\
  }\href {\doibase 10.1007/s00332-013-9186-1} {\bibfield  {journal} {\bibinfo
  {journal} {J. Nonlinear Sci.}\ }\textbf {\bibinfo {volume} {24}},\ \bibinfo
  {pages} {117} (\bibinfo {year} {2014})}\BibitemShut {NoStop}%
\bibitem [{\citenamefont {Goh}\ and\ \citenamefont {Scheel}(2015)}]{goh2}%
  \BibitemOpen
  \bibfield  {author} {\bibinfo {author} {\bibfnamefont {R.}~\bibnamefont
  {Goh}}\ and\ \bibinfo {author} {\bibfnamefont {A.}~\bibnamefont {Scheel}},\
  }\href@noop {} {\  (\bibinfo {year} {2015})},\ \Eprint
  {http://arxiv.org/abs/1512.08601} {arXiv:1512.08601} \BibitemShut {NoStop}%
\bibitem [{\citenamefont {Morrissey}\ and\ \citenamefont {Scheel}(2015)}]{mor}%
  \BibitemOpen
  \bibfield  {author} {\bibinfo {author} {\bibfnamefont {D.}~\bibnamefont
  {Morrissey}}\ and\ \bibinfo {author} {\bibfnamefont {A.}~\bibnamefont
  {Scheel}},\ }\href {\doibase 10.1137/15M1012554} {\bibfield  {journal}
  {\bibinfo  {journal} {SIAM J. Appl. Dyn. Syst.}\ }\textbf {\bibinfo {volume}
  {14}},\ \bibinfo {pages} {1387} (\bibinfo {year} {2015})}\BibitemShut
  {NoStop}%
\bibitem [{Note1()}]{Note1}%
  \BibitemOpen
  \bibinfo {note} {Note that only wavenumber-selecting strain-displacement
  relations $k=K(\varphi )$ yield well-posed mixed boundary conditions \cite
  {mor}.}\BibitemShut {Stop}%
\bibitem [{\citenamefont {Olmstead}\ and\ \citenamefont
  {Handelsman}(1976)}]{olmstead}%
  \BibitemOpen
  \bibfield  {author} {\bibinfo {author} {\bibfnamefont {W.~E.}\ \bibnamefont
  {Olmstead}}\ and\ \bibinfo {author} {\bibfnamefont {R.~A.}\ \bibnamefont
  {Handelsman}},\ }\href@noop {} {\bibfield  {journal} {\bibinfo  {journal}
  {SIAM Rev.}\ }\textbf {\bibinfo {volume} {18}},\ \bibinfo {pages} {275}
  (\bibinfo {year} {1976})}\BibitemShut {NoStop}%
\bibitem [{Note2()}]{Note2}%
  \BibitemOpen
  \bibinfo {note} {In direct simulations, for CGL, SH, and RD, we used domain
  sizes $L= 500\protect \ldots 2000$, for CH $L=2000 \protect \ldots 8000$. We
  used second order finite differences with $dx=0.005\protect \ldots 0.1$ for
  spatial discretization, and Matlab's ODE15s for time stepping.}\BibitemShut
  {Stop}%
\end{thebibliography}%
\end{document}